\documentclass[final]{l4dc2025}

\usepackage{xurl}

\newtheorem{assumption}{Assumption}
\usepackage{enumitem}
\usepackage{booktabs}
\usepackage{makecell}
\usepackage{multirow}
\usepackage{float}
\usepackage{times}
\usepackage{bm}

\newcommand{\rplus}[0]{[0, \infty)}

\title[Neural operators for predictor feedback control of nonlinear systems]{Neural Operators for Predictor Feedback Control of\\ Nonlinear Delay Systems}

\coltauthor{\Name{Luke Bhan}\thanks{Equal contribution} \Email{lbhan@ucsd.edu} \\ 
\Name{Peijia Qin}\footnotemark[1]\Email{pqin@ucsd.edu} \\ 
\Name{Miroslav Krstic} \Email{mkrstic@ucsd.edu} \\ 
\Name{Yuanyuan Shi} \Email{yyshi@ucsd.edu} \\ 
\addr University of California, San Diego}

\begin{document}
\maketitle
\begin{abstract}    
Predictor feedback designs are critical for delay-compensating controllers in nonlinear systems. However, these designs are limited in practical applications as predictors cannot be directly implemented, but require numerical approximation schemes, which become computationally prohibitive when system dynamics are expensive to compute. To address this challenge, we recast the predictor design as an operator learning problem, and learn the predictor mapping via a neural operator. We prove the existence of an arbitrarily accurate neural operator approximation of the predictor operator. Under the approximated predictor, we achieve semiglobal practical stability of the closed-loop nonlinear delay system. The estimate is semiglobal in a unique sense — one can enlarge the set of initial states as desired, though this increases the difficulty of training a neural operator, which appears practically in the stability estimate. Furthermore, our analysis holds for any black-box predictor satisfying the universal approximation error bound. We demonstrate the approach by controlling a 5-link robotic manipulator with different neural operator models, achieving significant speedups compared to classic predictor feedback schemes while maintaining closed-loop stability.
\end{abstract}

\vspace{3pt}

\section{Introduction}
Modern applications such as telerobotics \citep{5333120, doi:10.1126/scirobotics.abf2756}, renewable energy-based power systems \citep{magnusson2020distributed}, biomedical devices \citep{Sharma2011-bx}, and unmanned vehicles \citep{WANG2020109200, MAZENC2022100664} often encounter sensing, computation, and communication delays that can critically degrade performance. To deploy functional real-world controllers, engineers must design proactive, delay-compensating control laws. Consequently, dynamical systems under delays have been a consistent focal point for control design since the introduction of the Smith predictor in 1957 \citep{smith1957closer, henson1994time, https://doi.org/10.1002/aic.690350914, 935057, 1272267, 1272269, doi:10.1137/040616383}.

In this work, we focus on predictor feedback controllers~\citep{1103023, 1102288}. Predictor feedback designs mitigate delays by deploying the delay-free control law not with the system's current state, but with a \emph{prediction of its future state}. This approach has been developed for both linear~\citep{zhouBook, JANKOVIC2010510, CACACE2017455} and nonlinear systems, including state-dependent delays~\citep{10.1115/1.4005278}, delay-adaptive control~\citep{6704718}, and multi-input delay control~\citep{7458842}. However, for nonlinear delay systems, explicit predictor mappings are rarely available for controller implementation. To address this, engineers approximate the predictor mapping with numerical schemes such as finite difference and successive approximations~\citep[Chapter 4]{iassonBook}, which become burdensome when the system dynamics are expensive to compute. 

\begin{figure}[t]
    \centering
    \vspace{-20pt}
    \includegraphics[width=\textwidth]{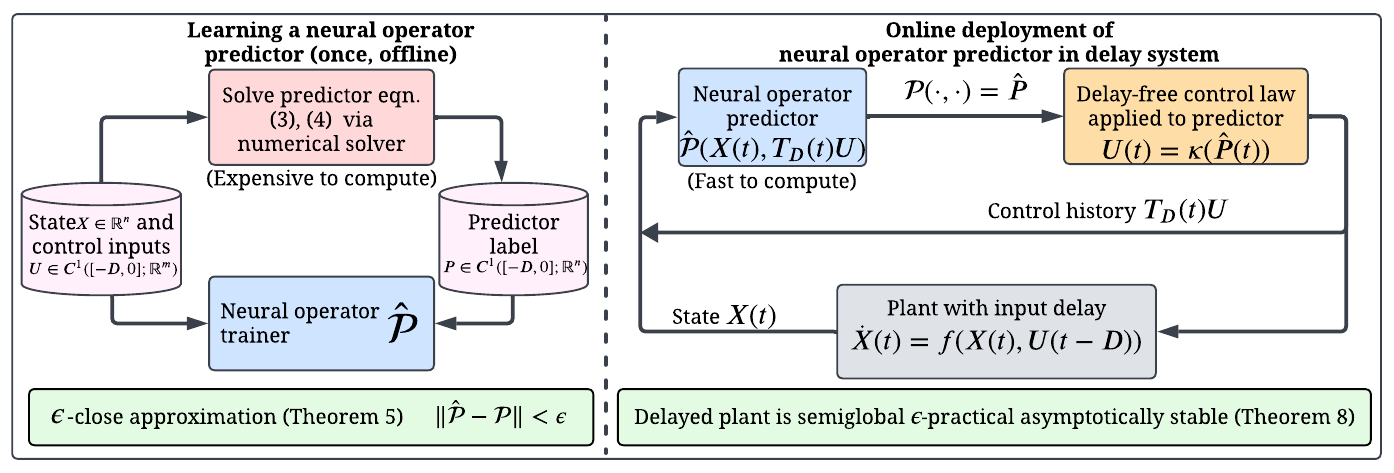}
    \vspace{-20pt}
    \caption{Overview of learning neural operator predictors for compensating delays.}
    \label{fig:arch}
    \vspace{-15pt}
\end{figure}
In this work, we introduce a new class of predictor feedback design with neural approximated predictors that are both \emph{theoretically stable} and \emph{computationally efficient}. Specifically, we recast the predictor mapping as a {mathematical operator} and propose learning a neural operator (NeuralOP) approximation of it (Figure \ref{fig:arch} left). Theoretically, we prove the continuity of the predictor operator, guaranteeing the existence of a NeuralOP that can approximate it to any desired accuracy. Computationally, we inherit the excellent speedups and scalability (with respect to discretization sizes) of NeuralOPs~\citep{li2021fourier, Lu2021}, highlighted across control designs including gain-scheduling~\citep{lamarque2024gainschedulingneuraloperator}, adaptive control~\citep{10374221, bhan2024adaptivecontrolreactiondiffusionpdes, lamarque2024adaptiveneuraloperatorbacksteppingcontrol}, traffic flows~\citep{pmlr-v242-zhang24c}, and PDE boundary control under delays~\citep{QI2024105714}.

To analyze the approximated predictor feedback design in closed-loop systems (Figure \ref{fig:arch} right), we present a {new analytical framework} for studying nonlinear delay systems with black-box predictor approximations. We show that under bounded predictor approximation errors, the resulting delay system achieves {semiglobal practical stability} (dependent on the error bound). Our analysis relies on a transport PDE representation of the delay which, coupled with an infinite-dimensional backstepping transformation, yields a perturbed system shown to be arbitrarily stable based on the approximation error. A key advantage of our analysis is that it applies to \emph{any black-box predictor} satisfying a universal approximation error bound. Thus, approximation techniques beyond NeuralOPs, such as Neural ODEs~\citep{NEURIPS2018_69386f6b}, recurrent neural networks~\citep{cho-etal-2014-properties,pmlr-v242-aguiar24a}, and generative models~\citep{CHEN2024112984}, are all covered under our analysis. 

Lastly, we demonstrate the proposed NeuralOP approximated predictor feedback control on a challenging 5-link robotic manipulator as in \citet{BAGHERI2019108485}, showcasing orders of magnitude speedups over numerical approximations in \citet{iassonBook} while maintaining stability.

\noindent \textbf{Notations.}
For functions, $f: [0, D] \times \mathbb{R} \to \mathbb{R}$, we use $f_x(x, t) = \frac{\partial f}{\partial x}(x, t)$ and $f_t(x, t) = \frac{\partial f}{\partial t}(x, t)$ to denote derivatives. $C^1([t-D, t];\mathbb{R}^m)$ denotes the set of functions with continuous first derivatives mapping from $[t-D, t]$ to $\mathbb{R}^m$. $T_D(t)f$ indicates the set $\{f(t+\theta); \theta \in [-D, 0]\}$ which represents the evaluation of the function $f$ with inputs from $[t-D, t]$. 
For a $n$-vector, we use $|\cdot|$ for the Euclidean norm. For functions, we define the spatial $L^p$ norms as $\|f(t)\|_{L^p[0, D]} = (\int_0^D |f(x, t)|^p dx)^{\frac{1}{p}}$ for $p\in [1, \infty)$. We use $\|f(t)\|_{L^\infty[0, D]} = \sup_{0 \leq x \leq D} |u(x, t)|$ to denote the spatial $L^\infty$ norm. 
\textcolor{blue}{Note: due to page limits, detailed proofs of the theoretical results and experimental results are provided in the Appendix of the extended online version \citep{bhan2024neural}.}

\section{Technical Background and Preliminaries}
In this work, we consider nonlinear systems with input delay of the form
\begin{eqnarray}
    \dot{X}(t) = f(X(t), U(t-D))\,, \label{eq:dynamics} 
\end{eqnarray}
where $X \in \mathbb{R}^n$ is the state vector, $U \in \mathbb{R}^m$ is the control, $f: \mathbb{R}^n \times \mathbb{R}^m \to \mathbb{R}^n$ represents the system dynamics, with $D \in \rplus$ is a constant, known delay. 

The classic predictor-feedback design consists of two components, a predictor $P(t)$, and a nominal, delay-free control law $\kappa(X(t))$. The delay-compensating control law $U(t)$ follows,
\begin{alignat}{1}
    U(t) &= \kappa(P(t))\,,  \label{eq:predictor-feedback-1} \\ 
    P(t) &= \int_{t-D}^t f(P(\theta), U(\theta)) d\theta + X(t)\,, \label{eq:predictor-feedback-2} 
\end{alignat}
where the initial condition for the integral equation is given by
\begin{equation}
    P(\theta) = \int_{-D}^\theta f(P(\sigma), U(\sigma)) d \sigma  +X(0)\,, \quad \theta \in [-D, 0)\,. \label{eq:predictor-feedback-ics}
\end{equation}
In essence, the predictor state $P(t)$, implicitly defined in \eqref{eq:predictor-feedback-2}, \eqref{eq:predictor-feedback-ics} is the $D$-second ahead prediction of the actual system state, i.e. $P(t) = X(t+D)$.

For stability analysis of the predictor feedback in nonlinear delays systems, both the classic design and the proposed neural predictor approximations, the following assumptions are needed. 
\begin{assumption} \label{assumption:forward-complete}
    $\dot{X} = f(X, \omega)$ in \eqref{eq:dynamics} is strongly forward complete.
\end{assumption}
\begin{assumption} \label{assumption:gas}
    There exists a control law $U(t) = \kappa(X(t))$ such that the delay-free system $\dot{X}(t) = f(X(t), U(t))$ is globally asymptotically stable. 
\end{assumption}
\begin{assumption}
\label{assumption:iss}
    The system $\dot{X} = f(X, \kappa(X) + \omega)$ in \eqref{eq:dynamics} is input-to-state stable (ISS). (See \citet{sontag1995characterizations} for ISS definition). 
\end{assumption}
All assumptions are standard in the predictor-feedback literature \citep{nikoBook}. 
Forward-completeness is needed to ensure the predictor system does not exhibit finite escape time and  we assume the existence of a globally asymptotically stabilizing control law for the system \emph{without delays}. Lastly, we use the ISS assumption to explicitly construct the Lyapunov functional.

\subsection{Exact predictor feedback designs for nonlinear delay systems}
For pedagogical organization as well as clarity, we now briefly review the stability analysis for the \emph{exact} control law \eqref{eq:predictor-feedback-1}, \eqref{eq:predictor-feedback-2}, \eqref{eq:predictor-feedback-ics} from  \citet{krsticDelay} highlighting the type of guarantees that we aim to attain under the proposed neural operator approximated predictor. We first present the stability result under the exact predictor feedback.

\begin{theorem} 
\label{thm:exact-predictor} \cite[Theorem 11]{krsticDelay} Let Assumptions \ref{assumption:forward-complete}, \ref{assumption:gas}, and \ref{assumption:iss} hold for \eqref{eq:dynamics}. Then, with the controller \eqref{eq:predictor-feedback-1}, \eqref{eq:predictor-feedback-2} for all $t\geq 0$, there exists $\beta_1 \in \mathcal{KL}$ such that
    \begin{equation}
         |X(t)| + \sup_{t-D \leq \tau \leq t}|U(\tau)| \leq \beta_1\left(|X(0)| + \sup_{-D \leq \tau \leq 0} |U(\tau)|, t\right)\,.
    \end{equation}    
\end{theorem}
To obtain Theorem \ref{thm:exact-predictor}, a crucial ingredient in the stability analysis 
is to represent the input delay in the system dynamics \eqref{eq:dynamics} as a {transport PDE}, coupled with the nonlinear ODE plant model,
\begin{subequations}
\begin{alignat}{2}
    \dot{X}(t) &= f(X(t), u(0, t))\,, \label{eq:transport-X}&& \quad t \in \rplus\,, \\
    u_t(x, t) &= u_x(x, t)\,, \label{eq:transport-dynamics} && \quad (x, t) \in [0, D) \times \rplus \,, \\ 
    u(D, t) &= \kappa(P(t)) \,, && \quad t \in \rplus\,. \label{eq:transport-BC}
\end{alignat}
\end{subequations}
However, the boundary condition in \eqref{eq:transport-BC} is unbounded and thus hinders the stability analysis of the coupled ODE-PDE system. To abate this issue, an infinite dimensional \emph{backstepping transformation} is employed to transform $u(x, t)$ into what we call the \emph{target system} $w(x, t)$, 
\begin{subequations}
    \begin{alignat}{2}
    \dot{X}(t) &= f(X(t), \kappa(X(t)) + w(0, t)) \,,  \label{eq:target-ode0}&& \quad  t \in \rplus\,,  \\
    w_t(x, t) &= w_x(x, t) \,, \label{eq:target-transport0}  && \quad (x, t) \in [0, D) \times \rplus \,,\\
    w(D, t) &= 0\,, \label{eq:target-bc0}&& \quad t \in \rplus\,.
\end{alignat}
\end{subequations}
The target system highlights the key advantage of the backstepping transformation as we now see the removal of the boundary condition in \eqref{eq:target-bc0} compared to \eqref{eq:transport-BC}. Thus, it is feasible to show that the $w$ transport PDE is globally exponentially stable in the $L^\infty$ norm (In fact, it is finite-time stable). From here,  one can combine the stability of the $w$-PDE with the ISS property of $X$, obtaining a bound on the coupled system. To complete the result of Theorem \ref{thm:exact-predictor}, one invokes the inverse backstepping transformation to transform the estimate in the $X$ and $w$ system back into the stability estimate on the $X$ and $u$ system. Interested readers could refer to \citet{krsticDelay} for details regarding the backstepping transformation and stability analysis of the exact predictor feedback.

While the exact predictor feedback ensures global asymptotic stability (Theorem \ref{thm:exact-predictor}), it relies on an explicit solution to the nonlinear system to obtain $P(t)$, which is rarely available and thus impractical to implement. Numerical methods like successive approximations and finite difference \citep{iassonBook} are often used instead but become computationally expensive for stiff systems or complex dynamics. To address these challenges, in this work, we will design neural operator based predictor feedback that is both implementable and preserves stability guarantees.

\subsection{Neural operators}
Neural operators (NeuralOP) are neural network approximations that extend the tradition notion of function approximation to nonlinear infinite-dimensional mappings of functions into functions. The excitement about NeuralOP came, in recent years, from their speed up of the solution of the notoriously hard nonlinear Navier-Stokes PDEs by {orders of magnitude}, while remaining discretization-invariant. The implementation of neural operators can take many forms, including DeepONet~\citep{Lu2021}, Fourier Neural Operator~\citep{li2021fourier}, NOMAD~\citep{seidman2022nomad}, and other NeuralOP architectures in domain-specific applications~\citep{shi2022machine,hao2023gnot,faroughi2024physics,azizzadenesheli2024neural}. 

A recent framework Nonlocal Neural Operator (NNO)~\citep{lanthaler2024nonlocalitynonlinearityimpliesuniversality} provides an abstract formulation that encompasses a majority of the popular of neural operator architectures including DeepONet and FNO. Under the NNO abstraction, the universal approximation theorem of NeuralOPs is given below. Let $\Omega_u \subset \mathbb{R}^{d_{u_1}}$, $\Omega_v \subset \mathbb{R}^{d_{v_1}}$ be bounded domains with Lipschitz boundary, and $\overline{\Omega_u}, \overline{\Omega_v}$ denote the closure of sets $\Omega_u, \Omega_v$ respectively. 
Let  $\mathcal{F}_c \subset C^0(\Omega_u; \mathbb{R}^c)$, $\mathcal{F}_v \subset C^0(\Omega_v; \mathbb{R}^v)$ be continuous function spaces. 

\begin{theorem}
    \label{thm:neural-operator-uat}
    \cite[Theorem 2.1]{lanthaler2024nonlocalitynonlinearityimpliesuniversality} Let $\Psi: C^0(\overline{\Omega_u};\mathbb{R}^{d_{u_1}}) \rightarrow C^0(\overline{\Omega_v}; \mathbb{R}^{d_{v_1}}) $ be a continuous operator and fix a compact set $K \subset C^0(\overline{\Omega_u};\mathbb{R}^{d_{u_1}})$. Then for any $\epsilon > 0$, there exists a single hidden layer neural operator $\hat{\Psi}: K \rightarrow C^0(\overline{\Omega_v}; \mathbb{R}^{d_{v_1}})$ satisfies  
\begin{equation}
    \sup_{u \in K} |\Psi(u)(y) - \hat{\Psi}(u)(y)|\leq \epsilon\,,
\end{equation}
for all values $y \in \Omega_v$.
\end{theorem}

The key corollary to Theorem~\ref{thm:neural-operator-uat} is that, many of the architectures aforementioned (e.g. FNO, DeepONet) are design instantiations of the NNO. Thus, they are universal operator approximators under the setting of Theorem \ref{thm:neural-operator-uat}.  
Henceforth, for the remainder of this study, we assume that any reference to a neural operator is an architecture of the NNO form. Interested readers could refer to \citet{lanthaler2024nonlocalitynonlinearityimpliesuniversality} for the detailed formalism of NNO and theoretical derivation of its universal approximation guarantees. 

\section{Stability under NeuralOP Approximated Predictor Feedback}
\label{sec:main_method}
We now present our main framework for delay-compensating feedback controllers under approximate predictors.  In Section \ref{sec:predictor_formulation}, we formulate the predictor approximation as an \emph{operator learning problem} and prove the existence of arbitrarily close NeuralOP based approximations to the predictor. Then, in Section \ref{sec:delay_perturbation}, we analyze the resulting target system $w$ under the approximate predictor. We develop a $L^\infty$ estimate of the \emph{perturbed} transport PDE where the perturbation is directly due to the approximation error. Lastly, we use the stability estimates of this perturbed PDE to analyze the coupled ODE-PDE system, proving semiglobal practical stability of the closed-loop system. Proofs of the theoretical results are referred to Appendix B of the online version~\cite{bhan2024neural}.


\subsection{Approximation of the predictor operator via NeuralOP}
\label{sec:predictor_formulation}
We begin by first defining the predictor operator as the solution to the ODE \eqref{eq:predictor-feedback-2}, \eqref{eq:predictor-feedback-ics} and proving existence of a uniform NeuralOP based approximation by showing contuity of the predictor operator. Naturally, we will let the predictor operator be the mapping from $X(t)$ and the history of the controls $T_D(t)U := \{U(t+\theta); \theta \in [-D, 0)\}$ into the solution to the predictor ODE \eqref{eq:predictor-feedback-2}, \eqref{eq:predictor-feedback-ics}.  
\begin{definition}(\textbf{Predictor operator})
\label{label}
Let $X \in \mathbb{R}^n$, $U\in C^1([-D, 0]; \mathbb{R}^m)$. Then, we define the predictor operator mapping as 
$\mathcal{P}: \left(X, U\right) \rightarrow P$ where $\mathcal{P}$ maps from  $\mathbb{R}^n\times C^1([-D, 0]; \mathbb{R}^m)$ to $ C^1([-D, 0]; \mathbb{R}^n)$ and  where $P(s) = \mathcal{P}(X, U)(s)$ satisfies for all $s\in [-D, 0]$,  
\begin{alignat}{1}
    \label{eq:predictor-operator-def} P(s) - \int_{-D}^s f(P(\theta), U(\theta)) d \theta - X = 0\,, \quad s \in [-D, 0]\,.
\end{alignat}    
\end{definition}

We aim to show that for any desired accuracy $\epsilon > 0$, there exists a NeuralOP approximation to the predictor operator $\mathcal{P}$. To do so, we require the assumption of Lipschitz dynamics. 
\begin{assumption} 
\label{assumption:lipschitz-dynamics}
Let $f(X, U)$ as in \eqref{eq:dynamics} and $\mathcal{X} \subset \mathbb{R}^n$ and $\mathcal{U} \subset \mathbb{R}^m$ be compact domains with bounds $\overline{\mathcal{X}}$ and $\overline{\mathcal{U}}$ respectively. Then, there exists a constant $C_f(\overline{\mathcal{X}}, \overline{\mathcal{U}})>0$ such that $f$ satisfies the Lipschitz condition
\begin{eqnarray}
    |f(x_1, u_1) - f(x_2, u_2)| \leq C_f(|x_1- x_2| + |u_1 - u_2|)\,, 
\end{eqnarray}
for all $x_1, x_2 \in \mathcal{X}$ and $u_1, u_2 \in \mathcal{U}$.
\end{assumption}
Assumption \ref{assumption:lipschitz-dynamics}, although strong, is necessary to ensure the continuity of the predictor and is similarly needed in numerical approximation schemes such as successive approximations \citep{iassonBook}. 
We first establish the continuity of the predictor $\mathcal{P}$ in the following Lemma. 
\begin{lemma} \label{lemma:operator-continuity}
    Let Assumption \ref{assumption:lipschitz-dynamics} hold. Then, for any $X_1, X_2 \in \mathbb{R}^n$ and control functions $U_1, U_2 \in C^1([-D, 0];\mathbb{R}^m)$, the predictor operator $\mathcal{P}$ defined in \eqref{eq:predictor-operator-def} satisfies 
    \begin{alignat}{1}
    \|\mathcal{P}(X_1, U_1) - \mathcal{P}(X_2, U_2)\|_{L^\infty([-D, 0])} \leq C_\mathcal{P} \left(|X_1-X_2| + \|U_1 - U_2\|_{L^\infty([-D, 0])}\right)\,,
    \end{alignat}
    with Lipschitz constant
    \begin{alignat}{1}
        C_\mathcal{P} := \max(1, DC_f)e^{DC_f}\,,
    \end{alignat}
    where $D$ is the delay constant and $C_f$ is the system Lipchitz constant defined in Assumption \ref{assumption:lipschitz-dynamics}.  
\end{lemma}


Applying the continuity of $\mathcal{P}$ (Lemma \ref{lemma:operator-continuity}) with the universal operator approximation Theorem \ref{thm:neural-operator-uat},  we establish the first main result on approximatibility of the predictor operator by neural operators. 
\begin{theorem} \label{thm:uat-predictor}
    Let $X \in \mathcal{X} \subset \mathbb{R}^n$ and $U \in C^1([-D, 0); \mathcal{U})$
    where $\mathcal{U} \subset \mathbb{R}^m$ is a bounded domain.
    Fix a compact set $K \subset \mathcal{X} \times C^1([-D, 0); \mathcal{U})$. Then, for all $\overline{\mathcal{X}}, \overline{\mathcal{U}}$,  $\epsilon > 0$, there exists a neural operator approximation $\hat{\mathcal{P}}: K \rightarrow C^1([-D, 0]; \mathbb{R}^n)$ such that 
    \begin{equation}
        \sup_{(X, U) \in K }|\mathcal{P}(X, U)(\theta) - \hat{\mathcal{P}}(X,U)(\theta)| < \epsilon\,, \quad \forall\theta \in [-D, 0]\,, 
    \end{equation}
    for all $X \in \mathcal{X}$, $U \in C^1([-D, 0]; \mathcal{U})$ such that $|X| < \overline{\mathcal{X}}$ and $\|U\|_{L^\infty([-D, 0])} < \overline{\mathcal{U}}$. 
\end{theorem}
First, notice that Theorem \ref{thm:uat-predictor} requires a compact domain of functions and thus restricts the possibility of \emph{global} stability guarantees. This type of compactness similarly appears in \citet[Section VI]{10374221} and is a common restriction of almost all universal approximation theorems. Further, Theorem \ref{thm:uat-predictor} guarantees the \emph{existence} of a neural operator approximation, but does not provide a minimum number of network parameters nor data samples to achieve such errors. To obtain practical estimates - which are architecture dependent - we refer the reader to \cite{10.1093/imatrm/tnac001} and \cite{10.5555/3546258.3546548}. These results can be directly applied in conjunction with Theorem \ref{thm:uat-predictor} to obtain an estimate of the minimal neural operator size to achieve a desired error bound $\epsilon$.

\subsection{Perturbed Transport PDE under the approximated predictor}
\label{sec:delay_perturbation}
Next, we analyze the closed-loop stability properties under the neural operator approximated predictor. For notational simplicity, define $\hat{P}$ as the solution to the operator approximation, namely
\begin{align}
    \hat{P}(t) = \hat{\mathcal{P}}(X(t), T_D(t)U)\,, \quad \forall t \geq 0\,.
\end{align}
Under the approximation, the system becomes 
\begin{subequations}
    \begin{alignat}{2}
    \dot{X}(t) &= f(X(t), u(0, t))\,, \label{eq:approx-transport-X} && \quad  t \in \rplus\,, \\
    u_t(x, t) &= u_x(x, t)\,, \label{eq:approx-transport-dynamics}  && \quad (x, t) \in [0, D) \times \rplus \,, \\ 
    u(D, t) &= \kappa(\hat{P}(t))\,,  && \quad t \in \rplus\,. \label{eq:appprox-transport-BC}
\end{alignat}
\end{subequations}
where \eqref{eq:appprox-transport-BC} differs from \eqref{eq:transport-BC} as we invoke the controller with the approximate predictor. Applying the following backstepping transformation,
\begin{alignat}{1}
    w(x, t) &= u(x, t) - \kappa(p(x, t))\,,  \label{eq:bcks-trans-forward} \\ 
    u(x, t) &= w(x, t) + \kappa(\pi(x, t))\,,\label{eq:bcks-trans-backward}
\end{alignat}
where $p(x, t)$ satisfies
\begin{alignat}{1}
\label{eq:predictor-pde-form}
    p(x, t) &= \int_0^x f(p(\xi, t), u(\xi, t)) d\xi +X(t)\,, \quad (x, t) \in [0, D] \times \rplus\,, 
\end{alignat} and $\pi(x, t)$ satisfies
\begin{eqnarray}
    \pi(x, t) = \int_0^x f(\pi(\xi, t), \kappa(\pi(\xi, t)) + w(\xi, t)) d\xi + X(t)\,, \label{eq:inverse-predictor-ode} 
\end{eqnarray}
we obtain the following target system.
\begin{lemma} 
\label{lemma:operator-target-system} 
The system \eqref{eq:approx-transport-dynamics}-\eqref{eq:appprox-transport-BC} under the backstepping transformation \eqref{eq:bcks-trans-forward}, \eqref{eq:bcks-trans-backward}
becomes,
\begin{subequations}
    \begin{alignat}{2}
    \dot{X}(t) &= f(X(t), \kappa(X(t)) + w(0, t))\,, \label{eq:target-approximate-ode} && \quad  t \in \rplus\,, \\ 
    w_t(x, t) &= w_x(x, t)\,, \label{eq:target-approximate-pde} && \quad (x, t) \in [0, D) \times \rplus \,, \\ 
    w(D, t) &= \underbrace{\kappa(P(t)) - \kappa(\hat{P}(t))} \label{eq:target-approximate-bc} \,, \quad && \quad t \in \rplus\,.
\end{alignat}
\end{subequations}
\end{lemma}
The key challenge is that \eqref{eq:target-approximate-bc} (compared to the exact predictor feedback in \eqref{eq:target-bc0}) contains a \emph{non-vanishing}, but arbitrarily small perturbation depending on the prediction error between $P$ and the operator approximation $\hat{P}$. As such, we introduce the following Lemma for characterizing the exact $L^\infty$ stability properties of the perturbed $w$ transport PDE under the predictor approximation error. 
\begin{lemma} \label{lemma:l-infty-target} (ISS estimate for $L^\infty$ norm of perturbed transport PDE) 
Let $c>0$ be a constant.
Then, the transport PDE \eqref{eq:target-approximate-pde}, \eqref{eq:target-approximate-bc} satisfies, for all $t \geq 0$, the following ISS-like stability estimate
\begin{eqnarray} \label{eq:semiglobalTargetPDE}
    \|w(t)\|_{L^\infty[0, D]} \leq e^{c(D-t)} \|w(0)\|_{L^\infty[0, D]} + e^{cD}\sup_{0 \leq s\leq t} |\kappa(P(t)) - \kappa(\hat{P}(t))| \,.
\end{eqnarray}
\end{lemma}
Notice that Lemma \ref{lemma:l-infty-target} does not guarantee asymptotic stability of $w$ to the equilibrium $0$, but instead to a set depending on the maximum difference between the predictor and its approximator which is uniformly bounded by $\epsilon$ - hence the system is globally asymptotically practically stable in $\epsilon$.

\subsection{Stability of the delay system under NeuralOP approximated predictor feedback}
\label{sec:stability_analysis}

Following Lemma \ref{lemma:l-infty-target}, we have now arrived at a bound on the target PDE \eqref{eq:target-approximate-pde}, \eqref{eq:target-approximate-bc} with boundary perturbation (underbraced) introduced by the predictor approximation error.
All that remains is to combine Lemma \ref{lemma:l-infty-target} with the coupled ODE system and the inverse backstepping transform \eqref{eq:bcks-trans-backward}. After a series of calculations detailed in Appendix \ref{appendix:main-result}, we arrive at the following main result. 

\begin{theorem}\label{thm:main-result}
    Let the system \eqref{eq:target-approximate-ode}, \eqref{eq:target-approximate-pde}, \eqref{eq:target-approximate-bc} satisfy Assumptions \ref{assumption:forward-complete}, \ref{assumption:iss}, \ref{assumption:lipschitz-dynamics} and assume that $\kappa$ is stabilizing as in Assumption \ref{assumption:gas}.    
    Then, for $\overline{\mathcal{B}}:= \min\{\overline{\mathcal{X}}, \overline{\mathcal{U}}\}$, there exists functions $\alpha_1 \in \mathcal{K}$ and $\beta_2 \in \mathcal{KL}$  such that if $\epsilon < \epsilon^*$ where 
    \begin{align}
    \epsilon^\ast(\overline{\mathcal{B}}) := \alpha_1^{-1}(\overline{\mathcal{B}})\,,
    \end{align}
    and the initial state is constrained to
    \begin{align}
    |X(0)| + \sup_{-D \leq \theta \leq 0}|U(\theta)| &\leq \Omega\,,
    \end{align}
    where
    \begin{align}
        \Omega(\epsilon, \overline{\mathcal{B}}) := \bar\alpha_2^{-1}\left(\overline{\mathcal{B}} - \alpha_1(\epsilon )\right)\,, \quad \bar\alpha_2(r) = \beta_2(r,0), \ \ r\geq 0\,,
    \end{align}
    then, the closed-loop solutions, under the controller with the neural operator-approximated predictor $U(t) = \kappa(\hat{P}(t))$, 
    satisfy the $\overline{\mathcal{B}}$-semiglobal and $\epsilon$-practical stability estimate 
     \begin{align}
        |X(t)| + \sup_{t-D \leq \theta \leq t}|U(\theta)| \leq \beta_2\left(|X(0)| + \sup_{-D \leq \theta \leq 0} |U(\theta)|, t\right) + \alpha_1(\epsilon)\,, \quad \forall t \geq 0\,, \label{eq:stability-main-res}
    \end{align}
    with the neural operator semiglobally trained with $\overline{\mathcal{B}} > \bar\alpha_2(\Omega)$ and $\epsilon \in (0, \alpha_1^{-1}(\overline{\mathcal{B}}))$.
\end{theorem}

Note, as aligned with the semiglobal property, the radius of initial conditions $\Omega$ can be made arbitrarily large given that one increases $\overline{\mathcal{B}}$. However, one will then pay a price as the neural operator will be harder to train to the desired $\epsilon$ for a larger set of $\overline{\mathcal{B}}$. Thus, practically, one will need a larger training set and more neurons in the architecture to achieve the same $\epsilon$ when increasing $\overline{\mathcal{B}}$. 
Additionally, it may be counter intuitive that $\epsilon^*$ grows with $\overline{\mathcal{B}}$, but the reader will recognize that a larger set of system states permits a larger transient and therefore a coarser approximation of the predictor. However, despite the larger tolerance in $\epsilon^*$, it is undesirable to have a large $\epsilon$ due to the fact the stability estimate is $\epsilon$-practical in \eqref{eq:stability-main-res}. Lastly, note in  the case of perfect approximation, i.e. $\epsilon=0$, we recover the stability estimate of the exact predictor feedback in Theorem \ref{thm:exact-predictor}. 

We briefly comment on the result of Theorem \ref{thm:main-result} compared to those of \citet[Theorem 5.1 and Theorem 5.3]{iassonBook}. Both results of \citet{iassonBook} obtain global asymptotic and local exponential stability, but require system dependent-assumptions on either the predictor implementation or the growth rate of the dynamics in order to annihilate the approximation error. In contrast, our result makes no assumption of the approximation scheme relying only on an uniform $\epsilon$ bound of the predictor error. Therefore, our framework encapsulates a larger class of approximators including black-box based predictor approximations with the trade-off of \emph{semiglobal practical stability} based on the approximation error $\epsilon$.    

\section{Numerical Results}
\label{sec:numerical_results}
In this section, we evaluate the performance of the proposed NeuralOP approximated predictor feedback via numerical experiments on a $n=5$ degree-of-freedom (DOF) robotic manipulator with a known, constant input delay. All data, models, and experiments are available on \href{https://github.com/lukebhan/NeuralOperatorPredictorFeedback/}{Github}
(\url{https://github.com/lukebhan/NeuralOperatorPredictorFeedback/}).
\color{black}

The dynamics of the 5 DOF robotic manipulator under input delay $D$ are governed by
\begin{align}
    M(X(t))\ddot{X}(t)+C(X(t), \dot{X}(t))\dot{X}(t)+G(X(t))=\tau(t-D), \quad t \in \rplus  \label{eq:baxter1}\,,
\end{align}
where $X(t)\in\mathbb{R}^n,\dot{X}(t)\in\mathbb{R}^n$, and $\ddot{X}(t)\in\mathbb{R}^n$ are the angles, angular velocities, and angular accelerations of the joints. Additionally, $\tau(t) \in\mathbb{R}^n$ indicates the vector of joint driving torques which is user-controlled. Lastly, $M(\cdot) \in \mathbb{R}^{n \times n}$, $C(\cdot, \cdot) \in \mathbb{R}^{n \times n}$, and $G(\cdot)\in \mathbb{R}^n$ are the mass, Coriolis, and gravitational matrices, which are symbolically derived using the Euler-Lagrange equations as in \cite{bagheri2017novel}.
We consider the trajectory tracking task of following a sinusoidal wave under an input delay $D=0.5$s. To develop the control law, we consider the feedback linearization approach of \cite{BAGHERI2019108485}.  More details about the experimental setup, dataset generation, and training are referred to Appendix C of the online version~\cite{bhan2024neural}.

\begin{table}[]
\centering
\caption{Performance of various predictor approximation approaches. 
In the right two columns, we consider the tracking task for $25$ samples beginning from randomized initial conditions and present both the tracking and prediction errors. }
\label{tab:baxter}
\resizebox{\textwidth}{!}{%
\begin{tabular}{l|c|c|c|c|c|c}
\hline
Approximation Method &
  \multicolumn{1}{l|}{Parameters} &
  \multicolumn{1}{l|}{\begin{tabular}[c]{@{}l@{}}Training error \\ $(L_2\ \text{error})$ $\downarrow$\end{tabular}} &
  \multicolumn{1}{l|}{\begin{tabular}[c]{@{}l@{}}Testing error \\ $(L_2\ \text{error})$ $\downarrow$\end{tabular}} &
  \multicolumn{1}{l|}{\begin{tabular}[c]{@{}l@{}}Avg. summed \\ trajectory tracking\\ error $(L_2\ \text{error})$ $\downarrow$\end{tabular}} &
  \multicolumn{1}{l|}{\begin{tabular}[c]{@{}l@{}}Avg. trajectory \\ prediction error \\ $(L_2\ \text{error})$ $\downarrow$\end{tabular}} &
  \multicolumn{1}{l}{\begin{tabular}[c]{@{}l@{}}Computation time\\ (1000 inst., ms.) $\downarrow$\end{tabular}} \\ \hline
Successive Approximations &
  --- &
  --- &
  --- &
  0.52 &
  $3.60 \times 10^{-7}$ &
  8.89 \\ \hline
FNO &
  594378 &
  {\color[HTML]{009901} \textbf{8.30$e{-5}$}} &
  {\color[HTML]{009901} \textbf{1.53$e{-4}$}} &
  {\color[HTML]{009901} \textbf{1.01}} &
  {\color[HTML]{009901} \textbf{0.15}} &
  {\color[HTML]{009901} \textbf{0.31}} \\
DeepONet &
  1641123 &
  3.00$e{-4}$ &
  4.16$e{-4}$ &
  3.01 &
  0.53 &
  {\color[HTML]{000000} 1.32} \\
FNO+GRU &
  158324 &
  1.84$e{-4}$ &
  3.02$e{-4}$ &
  3.50 &
  1.11 &
  0.45 \\
DeepONet+GRU &
  1756205 &
  1.63$e{-4}$ &
  3.26$e{-4}$ &
  1.58 &
  0.55 &
  1.42\\
  \hline
\end{tabular}%
}
\end{table}

For our predictor feedback approximations, we employ a series of approaches including the successive approximation approach in \cite{iassonBook}, classical neural operators (FNO \citep{li2021fourier} and DeepONet \citep{Lu2021}), and spatio-temporal neural operators (FNO + GRU and DeepONet + GRU \citep{michałowska2024neuraloperatorlearninglongtime}) that combine neural operators and recurrent neural networks. We choose to include the spatio-temporal neural operators as they have shown a strong propensity to handle time-correlated inputs as in the predictor case; however, we emphasize only FNO and DeepONet have the universal approximation guarantees of Theorem \ref{thm:uat-predictor}.

\begin{figure}[t]
    \centering
    \vspace{-10pt}
    \includegraphics[width=0.95\textwidth]{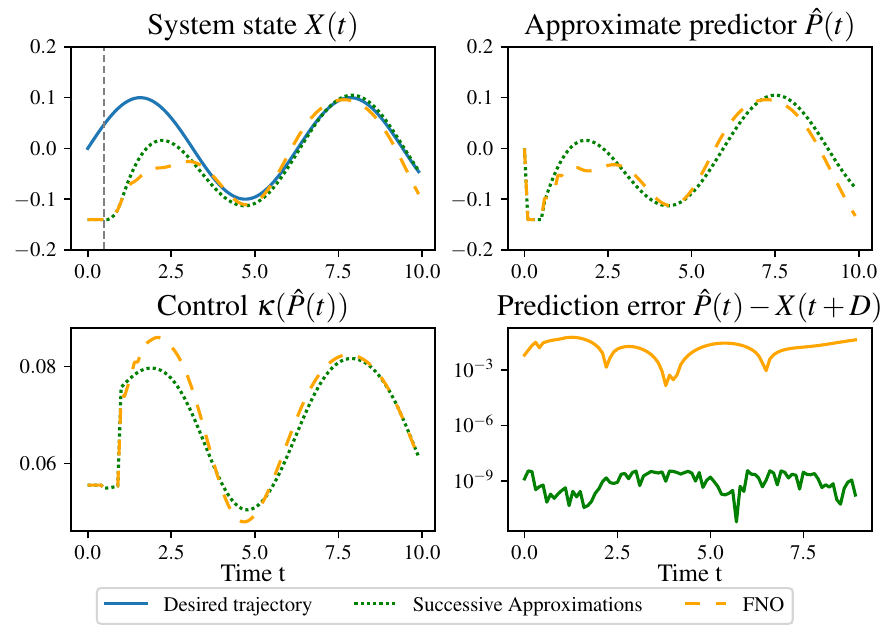}
    \vspace{-10pt}
    \caption{Example of a single trajectory of the robotic manipulator system. The fifth state (the last arm) and the fifth control input are displayed for clarity. The gray line in the top-left figure indicates the arrival of the first control signal at the delay of $D=0.5$ seconds. The bottom-right figure shows the prediction error under the NeuralOP predictor.}
    \label{fig:baxter}
    \vspace{-20pt}
\end{figure}

To develop a suitable dataset for learning, we simulate the robot dynamics with noisy perturbations to collect $250,000$ trajectory samples. In Table \ref{tab:baxter}, we present the training and testing $L_2$ errors. We see that each method is able to achieve an $L_2$ accuracy on the magnitude of $10^{-4}$. Furthermore, FNO outperforms the other approaches as the DeepONet methods struggle with training stabilization and the extra parameterization from the GRU leads to overfitting. Further, to evaluate the NeuralOP predictors in a predictor feedback controller, we simulate the closed-loop performance with the approximated predictors in $25$ trajectories with different initial conditions. 
For all $25$ trajectories, all methods stabilize (despite the less accurate approximation of the NeuralOP approach), and we report the summed tracking errors between $X(t+D)$ and $\hat{P}(t)$ in Table \ref{tab:baxter}. In the last column of Table \ref{tab:baxter}, we compare the average prediction time of the neural network based approaches with a \emph{single} iteration of the successive approximation method. For each method, we average over 1000 samples (Nvidia 4060) and as expected, the NeuralOP based approaches outperform the numerical scheme, achieving about $30\times$ speedups in the case of FNO.  Additionally, we emphasize that the neural operators scale extremely well with respect to both step size and delay length compared to the numerical predictor achieving speedups on the magnitude of $420\times$ presented in Appendix, under finer temporal discretizations. Lastly, we showcase an example trajectory in Figure \ref{fig:baxter} comparing the successive approximations approach with the FNO predictor. In Figure \ref{fig:baxter}, we see that, despite the coarse neural operator approximation, the state still maintains an $\epsilon$-ball around the target trajectory, which aligns with the $\epsilon$-practical stability estimate given in Theorem \ref{thm:main-result}. 

\section{Conclusion and Future Work}
In this work, we developed NeuralOP approximations of predictors and analyzed the resulting stability estimates in nonlinear delay systems. 
Particularly, (1) we defined the predictor operator and established its continuity, enabling the existence of an arbitrarily accurate NeuralOP approximation. Additionally, (2) we presented a semiglobal practical stability estimate whose initial conditions can be made arbitrarily large at the price of a more challenging NeuralOP approximation. This estimate is unique as it can be applied to \emph{any} black-box approximate predictor satisfying a uniform error bound. Lastly, (3) we highlighted the advantage of our approach on a 5-link robotic manipulator attaining orders of magnitude speedups compared to the numerical approximated predictor. 

Fundamentally, this is the first work to consider \emph{black-box} approximate predictors with uniform approximation errors. 
Our final result, Theorem \ref{thm:main-result}, presenting semiglobal practical stability for approximate predictors, is the first of its kind with minimal assumptions. It crucially relies on the universal approximation theorem of neural operators and the $L^\infty$ stability estimate of the transport PDE. Intentionally, this work focused on the simplest possible non-trivial problem - the constant, known input delay case - for which the analysis is  challenging. Additionally, the numerical experiments can be extended both in terms of delay independent operator designs as well as more challenging high-dimensional systems. Thus, by following the blueprint laid out in this work, there is ample opportunity for future research including problems featuring non-constant input delays, distributed input delays, and delay-adaptive predictor feedback designs with unknown input delays. 

\section*{Acknowledgments}
The work of Luke Bhan was supported
by the DOE under Grant DE-SC0024386. The work of Miroslav Krstic was supported in part by the AFOSR under Grant FA9550-231-0535 and in part by the NSF under Grant ECCS-2151525. The work of Yuanyuan Shi is supported by Department of Energy grant DE-SC0025495 and a Schmidt Sciences AI2050 Early Career Fellowship.

\bibliography{references}

\clearpage
\appendix

\section{Technical Background and Preliminaries}
\renewcommand{\thesubsection}{\Alph{section}.\arabic{subsection}}

\subsection{Strongly-forward complete} \label{appendix:strongly-forward-complete}
\begin{definition} \label{def:strongly-forward-complete} \cite[Definition 6.1]{krsticDelay} The system $\dot{X} = f(X, \omega)$ with $f(0, 0) = 0$ is \emph{strongly forward complete} if there exists a smooth function $R : \mathbb{R}^n \to \rplus$ and class $\mathcal{K}_{\infty}$ to decrease the merit function in each iteration, provided functions, $\alpha_1$, $\alpha_2$, $\alpha_3$ such that 
\begin{alignat}{1}
    \alpha_1(|X|) \leq R(X) \leq \alpha_2(|X|) \,, \\ 
    \frac{\partial R(X)}{\partial X}f(X, \omega) \leq R(X) + \alpha_3(|\omega|)\,,
\end{alignat}
for all $X \in \mathbb{R}^n$ and for all $\omega \in \mathbb{R}^m$. 
\end{definition}
\subsection{Input-to-state stability (ISS)} \label{appendix:iss}
\begin{definition} \citep{sontag1995characterizations} \label{def:iss}
    The system $\dot{X} = f(X, \kappa(X) + \omega)$ is \emph{input-to-state stable}(ISS) if there exists a class $\mathcal{KL}$ function $\beta$ and a class $\mathcal{K}$ function $\sigma$, such that for any $X(0)$, and for any continuous, bounded input $\omega(\cdot)$, the solution exists for all $t\geq 0$ and satisfies
    \begin{alignat}{1}
    |X(t)| \leq \beta(|X(t_0)|, t- t_0) + \sigma \left(\sup_{t_0 \leq \tau \leq t} |\omega(\tau)| \right)\,,
    \end{alignat}
    for all $t_0, t$ such that $0 \leq t_0 \leq t$. 
\end{definition}

\subsection{Review of predictor feedback for nonlinear delay systems} 
\label{appendix:predictor_feedback}
For pedagogical organization as well as clarity, we briefly review the stability analysis for the \emph{exact} control law \eqref{eq:predictor-feedback-1}, \eqref{eq:predictor-feedback-2}, \eqref{eq:predictor-feedback-ics} from  \citet{krsticDelay}. 
First, let rewrite the input delay in the system dynamics \eqref{eq:dynamics} as a {transport PDE}, coupled with the nonlinear ODE plant model,
\begin{subequations}
\begin{align*}
\dot{X}(t) &= f(X(t), u(0, t))\,, && \quad t \in \rplus\,, \\
    u_t(x, t) &= u_x(x, t)\,, && \quad (x, t) \in [0, D) \times \rplus \,, \\ 
    u(D, t) &= \kappa(P(t)) \,, && \quad t \in \rplus\,.
\end{align*}
\end{subequations}
where the predictor in PDE notation becomes $P(t) = p(D, t)$ with
\begin{align*}
    p(x, t) &= \int_0^x f(p(\xi, t), u(\xi, t)) d\xi +X(t)\,, \quad (x, t) \in [0, D] \times \rplus\,. 
\end{align*}
Notice that, although we have introduced a second variable $x$, the predictor $p(x,t)$ is equivalent to $P(t)$.
The core idea of rewriting \eqref{eq:dynamics} is that the delayed input $U(t-D)$ is now represented via the transport PDE allowing one to analyze stability of the coupled ODE-PDE system. However, the boundary condition of the transport PDE $u(D, t) = \kappa(P(t))$ is unbounded and thus hinders the stability analysis of the coupled system. 

To abate this issue, \cite{krsticDelay} introduces the following infinite dimensional \emph{backstepping transformation} which transforms $u(x, t)$ into what we call the \emph{target system} $w(x, t)$. 
\begin{align*}
    w(x, t) &= u(x, t) - \kappa(p(x, t))\,,   \\ 
    u(x, t) &= w(x, t) + \kappa(\pi(x, t))\,,
\end{align*}
where $\pi(x, t)$ satisfies
\begin{align*}
    \pi(x, t) = \int_0^x f(\pi(\xi, t), \kappa(\pi(\xi, t)) + w(\xi, t)) d\xi + X(t)
\end{align*}
First, note that, by definition $\pi(x, t) = p(x, t)$, but is rewritten in terms of the $w$ variables whereas $p(x, t)$ is written in terms of the $u$ variable. Second, note that the transformation between $w$ and $u$ is invertible, and therefore, analysis conducted on the $w$ system will correspondingly apply to the $u$ system. Third, under the backstepping transformation, we obtain the following target system,
\begin{subequations}
    \begin{align*}
    \dot{X}(t) &= f(X(t), \kappa(X(t)) + w(0, t)) \,,  && \quad  t \in \rplus\,,  \\
    w_t(x, t) &= w_x(x, t) \,, && \quad (x, t) \in [0, D) \times \rplus \,,\\
    w(D, t) &= 0\,, && \quad t \in \rplus\,.
\end{align*}
\end{subequations}
The target system highlights the key advantage of the backstepping transformation as we now see the removal of the boundary disturbance $w(D, t) = 0$. Thus, it is feasible to show that the $w$ transport PDE is globally exponentially stable in the $L^\infty$ norm (In fact, it is finite-time stable). From here,  one can combine the stability of the $w$ PDE with the ISS property of $X$, obtaining a bound on the coupled ODE-PDE system. To complete the result of Theorem \ref{thm:exact-predictor}, one needs to invoke the invertibility of the backstepping transformation to transform the estimate in the $X$ and $w$ system back into an estimate on the $X$ and $u$ system. 

Note that in our analysis of the predictor feedback under neural approximated predictor, the same backstepping transformation is employed. However, the target PDE we obtain differs from the canonical form due to the boundary perturbation introduced from the neural operator approximation error in \eqref{eq:target-approximate-bc}, which necessitates the ISS analysis of the perturbed transport PDE in Lemma \ref{lemma:l-infty-target}.  

\subsection{Review of non-local neural operators}
\label{appendix:nops}
Neural operators are finite dimensional approximations of nonlinear operators which map across function spaces. A neural operator takes a representation of the input function $c(x)$, its evaluation point $x$, and a target evaluation point $y$ which specifies the point to evaluate the target function after applying the operator to $c$. In practice, this infinite dimensional mapping is then approximated using finite dimensional neural networks. 

We now review the non-local neural operator abstraction introducted in \cite{lanthaler2024nonlocalitynonlinearityimpliesuniversality}, which unifies different neural operator designs from a rigorous analytical perspective. Namely, let $\Omega_u \subset \mathbb{R}^{d_{u_1}}$, $\Omega_v \subset \mathbb{R}^{d_{v_1}}$ be bounded domains with Lipschitz boundary and let  $\mathcal{F}_c \subset C^0(\Omega_u; \mathbb{R}^c)$, $\mathcal{F}_v \subset C^0(\Omega_v; \mathbb{R}^v)$ be continuous function spaces. Then, a non-local neural operator approximation for the nonlinear operator $\Psi$ as any function satisfying the following form:
\begin{definition}[Non-local Neural Operators] \label{definition:neural-operator} \cite[Section 1.2]{lanthaler2024nonlocalitynonlinearityimpliesuniversality}
    Given a channel dimension $d_c > 0$, we call any $\hat{\Psi}$ a neural operator given it satisfies the compositional form $\hat{\Psi} = \mathcal{Q} \circ \mathcal{L}_L \circ \cdots \circ \mathcal{L}_1 \circ \mathcal{R}$ where  $\mathcal{R}$ is a lifting layer, $\mathcal{L}_l, l=1,..., L$ are the hidden layers, and $\mathcal{Q}$ is a projection layer. That is, 
    $\mathcal{R}$ is given by 
    \begin{equation}
    \mathcal{R} : \mathcal{F}_c(\Omega_u; \mathbb{R}^c) \rightarrow \mathcal{F}_s(\Omega_s; \mathbb{R}^{d_c}), \quad c(x) \mapsto R(c(x), x)\,, 
\end{equation} where $\Omega_s \subset \mathbb{R}^{d_{s_1}}$, $\mathcal{F}_s(\Omega_s; \mathbb{R}^{d_c})$ is a Banach space for the hidden layers and $R: \mathbb{R}^c \times \Omega_u \rightarrow \mathbb{R}^{d_c}$ is a learnable neural network acting between finite-dimensional Euclidean spaces. For $l=1, ..., L$, each hidden layer is given by 
\begin{equation} \label{eq:generalNeuralOperator}
    (\mathcal{L}_l v)(x) := s \left( W_l v(x) + b_l + (\mathcal{K}_lv)(x)\right)\,, 
\end{equation}
where weights $W_l \in \mathbb{R}^{d_c \times d_c}$ and biases $b_l \in \mathbb{R}^{d_c}$ are learnable parameters, $s: \mathbb{R} \rightarrow \mathbb{R}$ is a smooth, infinitely differentiable activation function that acts component wise on inputs and $\mathcal{K}_l$ is the nonlocal operator given by 
\begin{equation} \label{eq:generalKernel}
    (\mathcal{K}_lv)(x) = \int_\mathcal{X} K_l(x, y) v(y) dy\,,
\end{equation}
where $K_l(x, y)$ is a kernel function containing learnable parameters. Lastly, the projection layer $\mathcal{Q}$ is given by 
\begin{equation}
    \mathcal{Q} : \mathcal{F}_s(\Omega_s; \mathbb{R}^{d_c}) \rightarrow \mathcal{F}_v(\Omega_v; \mathbb{R}^v), \quad s(x) \mapsto Q(s(x), y)\,, 
\end{equation}
where $Q$ is a finite dimensional neural network from $\mathbb{R}^{d_c} \times \Omega_v \rightarrow \mathbb{R}^v$. 
\end{definition}

This abstract formulation of neural operators covers a wide range of architectures including the Fourier Neural Operator (FNO) \cite{li2021fourier} as well as DeepONet \cite{Lu2021} where the differentiator between these approaches lies in the implementation of the kernel function $K_l$. However, in \cite{lanthaler2024nonlocalitynonlinearityimpliesuniversality}, the authors showed that a single hidden layer neural operator with a kernel given by the averaging kernel $K_l(x, y) =1/|\mathcal{X}|$ where $|\mathcal{X}|$ is the diameter of the domain is sufficient for universal operator approximation, as stated in Theorem \ref{thm:neural-operator-uat}.  

\section{Proofs of Technical Results in Section \ref{sec:main_method}}
\renewcommand{\thesubsection}{\Alph{section}.\arabic{subsection}}
\subsection{Proof of Lemma \ref{lemma:operator-continuity}}
For all $s \in [-D, 0]$, let $P_1(s) := \mathcal{P}(X_1, U_1)(s)$ and likewise $P_2(s) := \mathcal{P}(X_2, U_2)(s)$. Then, we have 
\begin{alignat}{1}
    P_1(s) - P_2(s) &= X_1 - X_2 + \int_{-D}^s f(P_1(\theta), U_1(\theta)) - f(P_2(\theta), U_2(\theta)) d \theta   \nonumber \\ 
    & \leq |X_1-X_2| + \int_{-D}^s C_f(|P_1(\theta) - P_2(\theta)| + |U_1(\theta) - U_2(\theta)|) d \theta \nonumber  \\
    &\leq  |X_1-X_2| + DC_f\|U_1-U_2\|_{L^\infty([-D, s])} + \int_{-D}^s C_f(|P_1(\theta) - P_2(\theta)|) d \theta \nonumber \\ 
    &\leq \exp(DC_f) (|X_1-X_2| + DC_f (\|U_1-U_2\|_{L^\infty([-D, s])}) \,,
\end{alignat}
where we used the definition of the predictor, Lipschitz continuity in Assumption \ref{assumption:lipschitz-dynamics}, properties of the $L^\infty$ norm in the integral, and Gronwall's inequality. 
Now, noting that the bound is monotonically increasing with respect to $s$, we can take $s \to 0$ to yield the result. 
\subsection{Proof of Theorem \ref{thm:uat-predictor}}
Note that $\mathcal{X}$ and $\mathcal{U}$ are both bounded domains. Further, from Lemma \ref{lemma:operator-continuity}, the operator $\mathcal{P}$ is continuous. Thus, we apply Theorem  \ref{thm:neural-operator-uat} to achieve the result. 
\subsection{Proof of Lemma \ref{lemma:operator-target-system}}
    The PDE representation of the system with the approximate predictor is as follows
    \begin{alignat}{2}
        \dot{X}(t) &= f(X(t), u(t, 0))\,, \quad && t \in \rplus  \\ 
        u_t(x, t) &= u_x(x, t)\,,\quad  && (x, t) \in [0, D) \times \rplus \,, \\ 
        u(D, t) &= \kappa(\hat{P}(t))\,, \quad && t \in \rplus\,.
    \end{alignat}
    Under the transform with the exact predictor, namely
    \begin{alignat}{1} \label{eq:bcks-tran-what-exact}
        w(x, t) = u(x, t) - \kappa(p(x, t))
    \end{alignat}
    we have 
    \begin{alignat}{1}
        w_t &= u_t - \frac{\partial}{\partial t} k(p(x, t)) \\ 
        w_x &= u_x - \frac{\partial}{\partial x} k(p(x, t))
    \end{alignat}
    Noting, that $u_t = u_x$ and noting that the function $p(x, t)$ is really a function of only one variable, namely $x+t$ as it is the solution to the ODE \eqref{eq:predictor-feedback-2}, yields $w_t = w_x$. Substituting the boundary conditions at $x=D$ into \eqref{eq:bcks-tran-what-exact} and noting $u(D, t) = \kappa(\hat{P}(t))$ yields \eqref{eq:target-approximate-bc}. 
\subsection{Proof of Lemma \ref{lemma:l-infty-target}}
As standard with transport PDEs, we aim to analyze the exponentially weighted norm of the form 
\begin{align}
     \|w(t)\|_{c,\infty} = \sup_{x \in [0, D]}|e^{cx}w(x, t)| = \lim_{p \to \infty} \left( \int_0^D e^{pcx}|w(x, t)|^p dx\right)^{\frac{1}{p}} \label{eq:exponentialNorm}\,, 
\end{align}
where $c>0$ which we note satisfies
\begin{eqnarray}
    \|w(t)\|_{L^\infty[0, D]} \leq \|w(t)\|_{c, L^\infty[0, D]} \leq e^{cD} \|w(t)\|_{L^\infty[0, D]}\,.
\end{eqnarray}
Thus, we will begin by analyzing the Lyapunov functional of $L^p$ norm, namely for $p \in (1, \infty)$, we have
\begin{align}
    V(t) = \|w(t)\|_{c, p}^p\,.
\end{align}
Taking the derivative, substitution of $w_t = w_x$ and applying integration by parts yields
\begin{align}
     \dot{V}(t)  &=  p\int_0^D e^{pcx}  \text{sgn}(w(x, t)) |w(x, t)|^{p-1} w_t(x, t) dx \nonumber \,, \quad &&\text{Differentiation}  \\ 
   &=  p \int_0^D e^{pcx} \text{sgn}(w(x, t)) |w(x, t)|^{p-1} w_x(x, t) dx\,,\quad &&\text{Substitution $w_t=w_x$} \nonumber  \\ 
   &=  e^{pcx} |w(x, t)|^p|_0^D -pc \int_0^D e^{pcx} |w(x, t)|^p dx\,,\quad &&\text{Integration by parts} \nonumber \\ 
   &=e^{pcD}|w(D, t)|^p - |w(0, t)|^p - pcV \nonumber\,, \quad &&\text{Substitution of $V$} \\ 
   &\leq e^{pcD}|w(D, t)|^p - pcV \,. \label{eq:lyapunovBound}\quad && \text{Properties of inequality}
\end{align}
From \eqref{eq:lyapunovBound}, we obtain
\begin{align}
      V(t) &\leq V(0) e^{-pct} + \frac{e^{pcD}}{pc} \sup_{0 \leq \tau \leq t}|w(D, \tau)|^p\,. \label{eq:lp-bound}
\end{align}
Now, substituting $V(t)$ into \eqref{eq:exponentialNorm} and noting that $ \lim_{p \to \infty} \|w(0)\|_{c, L^p[0, D]} = \|w(0)\|_{c, L^\infty[0, D]}$ yields 
\begin{align}
    \|w(t)\|_{c, \infty} &= \lim_{p \to \infty} \left( V(t) \right)^\frac{1}{p} \nonumber \\ 
    &\leq \lim_{p \to \infty} \left( \|w(0)\|_{c, L^p[0, D]} e^{-ct} + e^{cD}\left(\frac{1}{cp}\right)^\frac{1}{p} \sup_{0 \leq \tau \leq t}|w(D, \tau)| \right)\nonumber  \\ 
    &\leq \|w(0)\|_{c, L^\infty[0, D]} e^{-ct} + e^{cD} \sup_{0 \leq \tau \leq t}|w(D, \tau)|
\end{align}
Noting that the weighted norm satisfies
        \begin{align}
            \|w(t)\|_{L^\infty[0, D]} &\leq \|w(t)\|_{c, L^\infty[0, D] } \nonumber \\  &\leq \|w_0\|_{c, L^\infty[0, D]} e^{-ct} + e^{cD} \sup_{0 \leq \tau \leq t} |w(D, \tau)| \nonumber \\ &\leq  \|w_0\|_{L^\infty[0, D]} e^{c(D - t) } + e^{cD} \sup_{0 \leq \tau \leq t} |w(D, \tau)|\,, \label{eq:infinityToregInequalities}
        \end{align}
        yields the final result. 
        
        The reader familiar with ISS estimates will note that this estimate is not of the typical form in that we do not explicitly achieve a bound on $\frac{d \|w(t)\|_{c, \infty}}{dt}$ and invoke the comparison Lemma directly. Instead, we achieve such a result on $\|w(t)\|_{c, p}^p$ and take the limit. This approach is common in supremum norm estimates of the transport PDE \citep{KARAFYLLIS2020104594, issBook, BASTIN2021112300} and the main advantage over \citet{krsticDelay} is avoiding the complications with negative exponents which are undefined at $w(t) \equiv 0$. 
    \subsection{Proof of Theorem \ref{thm:main-result}} \label{appendix:main-result}
    We begin by bounding the function $|X(t)| + \|w(t)\|_{L^\infty[0, D]}$. First, applying Assumption \ref{assumption:iss}, we know there exists class $\mathcal{K}_\infty$ functions $\alpha_2$, $\alpha_3$, $\alpha_4$, $\alpha_5$ such that 
    \begin{align}
        \alpha_2(|X(t)|) &\leq S(X(t)) \leq \alpha_3(|X(t))\,, \label{eq:iss-1} \\
        \frac{\partial S(X(t))}{\partial Z} f(X(t), \kappa(X(t))+w(0, t)) &\leq -\alpha_4(|X(t)|) + \alpha_5(|w(0, t)|)\,.
    \end{align}
    Now, we first bound $|X(t)|$. We begin by defining the Lyapunov function
    \begin{align}
       \breve{V}(t) = S(X(t)) \,,
    \end{align}
    as in \eqref{eq:iss-1}. Then, we have that 
    \begin{align}
        \dot{\breve{V}} &\leq -\alpha_4(|X(t)|) + \alpha_5(|w(0, t)|)  \nonumber \\
        &\leq -\alpha_4(|X(t)|) +  \alpha_5\left(\sup_{0 \leq x \leq D}|w(x, t)|\right) \nonumber  \nonumber \\
        &=  -\alpha_4(|X(t)|) + \alpha_5\left(\|w(t)\|_{L^\infty[0, D]}\right)\nonumber \\ 
        &\leq -\alpha_4(|X(t)|) + 
        \alpha_5\left(\|w(0)\|_{L^\infty[0, D]}e^{c(D-t)} + e^{cD}\sup_{0 \leq \tau \leq t}|w(D, \tau)|\right)\,, 
    \end{align}
    where in the last inequality, we applied the stability bound in Lemma \ref{lemma:l-infty-target}. 
    Reproducing the argument from \citet[Lemma C.4]{kkk}, we have that there exists $\beta_3 \in \mathcal{KL}$, $\alpha_7 \in \mathcal{K}_\infty$ such that
    \begin{eqnarray}
        \breve{V}(t) \leq \beta_3(|X(0)|, t) +\alpha_7\left(\|w(0)\|_{L^\infty[0, D]}e^{c(D-t)} + e^{cD}\sup_{0 \leq \tau \leq t}|w(D, \tau)|\right) \,.
    \end{eqnarray}
    Now applying \eqref{eq:iss-1}, we have, there exists $\beta_4 \in \mathcal{KL}$, $\alpha_8 \in \mathcal{K}_\infty$ such that 
    \begin{eqnarray}
        |X(t)| \leq \beta_4(|X(0)|, t) +\alpha_8\left(\|w(0)\|_{L^\infty[0, D}e^{c(D-t)} + e^{cD}\sup_{0 \leq \tau \leq t}|w(D, \tau)|\right)\,.
    \end{eqnarray}
    Now, using the triangle inequality for class $\mathcal{K}$ functions, ($\alpha_8(x+y) \leq \alpha_8(2x) + \alpha_8(2y)$), we have that
    \begin{align}
        |X(t)| &\leq \beta_4(|X(0)|, t) + \alpha_8\left(2\|w(0)\|_{L^\infty[0, D]}e^{c(D-t)}\right) + \alpha_8\left(2e^{cD}\sup_{0 \leq \tau \leq t}|w(D, \tau)|\right)\,.
    \end{align}
    It is clear that the middle term is decreasing with respect $t$ and thus, we have that there exists some function $\beta_5 \in \mathcal{KL}$ such that 
    \begin{align}
        |X(t)| &\leq \beta_5\left(|X(0)| + \|w(0)\|_{L^\infty[0, D]}, t\right) + \alpha_8\left(2e^{cD}\sup_{0 \leq \tau \leq t}|w(D, \tau)|\right)\,. \label{eq:x-estimate}
    \end{align}
    Now, combining the estimate \eqref{eq:x-estimate} with the estimate on $w$ in Lemma \ref{lemma:l-infty-target}, we have
    \begin{align}
        |X(t)| + \|w(t)\|_{L^\infty[0, D]} &\leq&& \beta_5\left(|X(0)| + \|w(0)\|_{L^\infty[0, D]}, t\right) + \alpha_8\left(2e^{cD}\sup_{0 \leq \tau \leq t}|w(D, \tau)|\right) \nonumber \\ & &&  + \|w(0)\|_{L^\infty[0, D]} e^{c(D - t) } + e^{cD} \sup_{0 \leq \tau \leq t} |w(D, \tau)| \label{eq:both-estimate}\,,
    \end{align}
    Note that \eqref{eq:both-estimate} implies the existence of $\beta_6 \in \mathcal{KL}$ and $\alpha_9 \in \mathcal{K}$ such that
    \begin{align}
        |X(t)| + \|w(t)\|_{L^\infty[0, D]} & \leq \beta_6\left(|X(0)| + \|w(0)\|_{L^\infty[0, D]}, t\right) + \alpha_9\left (\sup_{0 \leq \tau \leq t}|w(D, \tau)|\right)\,.
    \end{align} 
    Lastly, we substitute in for $\sup_{0 \leq \tau \leq t} |w(D, \tau)|$ yielding
     \begin{align}
        |X(t)| + \|w(t)\|_{L^\infty[0, D]} & \leq \beta_6\left(|X(0)| + \|w(0)\|_{ L^\infty[0, D]}, t\right) + \alpha_9\left (\sup_{0 \leq \tau \leq t}|\kappa(P(t)) - \kappa(\hat{P}(t))|\right)\,.\label{eq:Xwfinalbound}
    \end{align} 
    To complete the proof, notice we used the exact backstepping transformation in \eqref{eq:bcks-trans-forward}, \eqref{eq:bcks-trans-backward} and thus can reapply the following two technical Lemmas from \citet{krsticDelay} 
    \begin{lemma} \label{lemma:Xwbound} \cite[Lemma 8]{krsticDelay}
        Let \eqref{eq:predictor-feedback-2} satisfy Assumption \ref{assumption:forward-complete} and consider \eqref{eq:bcks-trans-forward} as its output map. Then, there exists $\alpha_{10} \in \mathcal{K}_\infty$ such that 
        \begin{eqnarray} 
            |X(t)| + \|w(t)\|_{L^{\infty}[0, D]} \leq \alpha_{10}\left(|X(t)| + \|u(t)\|_{L^\infty[0, D]}\right)\,. 
        \end{eqnarray}
    \end{lemma}
    \begin{lemma} \label{lemma:Xubound} \cite[Lemma 9]{krsticDelay}
        Let \eqref{eq:inverse-predictor-ode} satisfy Assumption \ref{assumption:iss} and consider \eqref{eq:bcks-trans-backward} as its output map. Then there exists $\alpha_{11} \in \mathcal{K}_\infty$ such that 
        \begin{eqnarray}
            |X(t)| + \|u(t)\|_{L^{\infty}[0, D]} \leq \alpha_{11} \left(|X(t)| + \|w(t)\|_{L^{\infty}[0, D]}\right)\,. 
        \end{eqnarray}
    \end{lemma}
    Combining Lemma \ref{lemma:Xwbound}, Lemma, \ref{lemma:Xubound} and \eqref{eq:Xwfinalbound} yields
    \begin{alignat}{2}
        |X(t)| + \|u(t)\|_{L^\infty[0, D]} & \leq&& \alpha_{11}(|X(t)|+\|w(t)\|_{L^\infty[0, D]} \nonumber  \\ 
        &\leq&& \alpha_{11}\left(\beta_6(|X(0)| + \|w(0)\|_{L^\infty[0, D]}, t) + \alpha_9\left (\sup_{0 \leq \tau \leq t}|\kappa(P(t)) - \kappa(\hat{P}(t))|\right)\right) \nonumber  \\
        &\leq&& \alpha_{11}\bigg(\beta_6 \left(\alpha_{10}\left( |X(0)| + \|u(0)\|_{L^\infty[0, D]}\right), t\right) \nonumber \\ & && + \alpha_9\left (\sup_{0 \leq \tau \leq t}|\kappa(P(t)) - \kappa(\hat{P}(t))|\right) \bigg)\,,
    \end{alignat}
    Applying properties of class $\mathcal{K}$ functions implies the existence of $\beta_2 \in \mathcal{KL}$, and $\alpha_{12} \in \mathcal{K}_\infty$ such that 
    \begin{align}
        |X(t)| + \|u(t)\|_{L^\infty[0, D]} \leq \beta_2 (|X(0)| + \|u(0)\|_{L^\infty[0, D]}, t) + \alpha_{12}\left (\sup_{0 \leq s \leq t} |\kappa(P(t)) - \kappa(\hat{P}(t))|\right)\,.
    \end{align}
    Now, noting that $\kappa$ is continuous (Assumption \ref{assumption:gas}), there exists a class $\mathcal{K}_\infty$ function $\alpha_{13}$
    such that $\sup_{0 \leq s \leq t}|\kappa(P(t)) - \kappa(\hat{P}(t))| \leq \alpha_{13}(\epsilon)$ where $\epsilon$ is as in Theorem \ref{thm:uat-predictor}. 
    The result then follows by letting $\alpha_1 = \alpha_{12} \circ \alpha_{13}$ and returning back to ODE notation. 

\section{Details for Numerical Experiment in Section \ref{sec:numerical_results}} \label{appendix:experimentDetails}
\subsection{Derivation of feedback controller for robotic manipulators}
\label{appendix:baxter}
Consider the $5$-link robot manipulator  
with the mathematical modeling given in \citet{BAGHERI2019108485}
\begin{align}
    M(X)\ddot{X}+C(X,\dot{X})\dot{X}+G(X)=\tau \label{eq:baxter-appendix}\,, 
\end{align}
where $X\in\mathbb{R}^5,\dot{X}\in\mathbb{R}^5$, and $\ddot{X}\in\mathbb{R}^5$ are the angles, angular velocities, and angular accelerations of the joints. $\tau\in\mathbb{R}^5$ indicates the vector of joint driving torques which is user controlled. 
Lastly, $M(X)\in\mathbb{R}^{5\times5}$, $C(X,\dot{X})\in\mathbb{R}^{5\times5}$, and $\dot G(X)\in\mathbb{R}^5$ are the mass, Coriolis, and gravitational matrices which are symbolically derived using the Euler-Lagrange equations \citep{bagheri2017novel}.
The multi-input nonlinear system (\ref{eq:baxter1}) can be written as $10^{\mathrm{th}}$-order ODE with the following general state-space form, 
\begin{align}
   \dot{\breve{X}}=f_0(\breve{X},U) \label{eq:baxter2}\,, 
\end{align}
where $  \breve{X}=[X_1,\cdots,X_5,\dot{X}_1,\cdots,\dot{X}_5]^T\in\mathbb{R}^{10}$ is the vector of states
and $U = \tau \in\mathbb{R}^{5}$ is the input of nonlinear system (\ref{eq:baxter2}).
Since we are working on the trajectory tracking task of following $X_{\text{des}} \in \mathbb{R}^5$, we consider the error dynamics of the form 
\begin{align}
    \dot{E} = f(E, U)\,, 
\end{align}
where $E=[e_1,e_2]^T\in\mathbb{R}^{10}$ is the vector of error states defined by 
\begin{align}
e_1&=X_{\mathrm{des}}-X\,, \\
e_2&=\dot{e}_1+\alpha e_1\,, 
\end{align}
where $\alpha\in\mathbb{R}^{7\times7}$ is a constant positive definite matrix.
Then, following \citet{BAGHERI2019108485}, the error dynamics become 
\begin{align}
    \begin{bmatrix}\dot{e_1}\\\dot{e_2}\end{bmatrix}=
    \begin{bmatrix}e_2-\alpha e_1\\\alpha e_2+h-M^{-1}\tau\end{bmatrix}\,, 
\end{align}
where $h=\ddot{X}_\text{des}-\alpha^2e_1+M^{-1}(C\dot{X}_\text{des}+G+C\alpha e_1-Ce_2)$. 
The feedback linearization based control law $\tau$ is derived as
\begin{align}
    \tau=\kappa(E)=M(h+(\beta+\alpha)e_2)\,,
\end{align}
with $\beta \in \mathbb{R}^{5 \times 5}$ is any positive definite matrix. We refer the readers to \citet{BAGHERI2019108485} for more details.

\subsection{Experimental settings}
\label{appendix:experiments}

\textbf{\underline{Manipulator parameters and trajectory sampling:}}
Following the system dynamics and control law in Appendix \ref{appendix:baxter}, we set $\alpha=\beta=I_{5}$, where $I$ is the identity matrix. Based off the Baxter robot implementation in \cite{7558740}, we define the maximum and minimum joint angles for each arm (radians) as 
\begin{align*}
\text{max\_joint\_lim}&=\{1.7016, 1.047, 3.0541, 2.618, 3.059\}\,, \\\text{min\_joint\_lim}&=[-1.7016, -2.147, -3.0541, -0.05, -3.059]\,,
\end{align*}
and chose the trajectory to be tracked as
\begin{align*}
X_\text{des}(t) = 0.1[\sin(t),\sin(t), \sin(t), \sin(t), \sin(t)] + (\text{max\_joint\_limit + } \text{min\_joint\_limit)/2.0}\,,
\end{align*}
which is a sinusoidal wave centered at the middle of each joint limit. Furthermore, in our implementation, we ensure that the torques are always saturated such that the control is always between 
\begin{align*}
          \text{max\_torque\_lim}=&\{50, 50, 50, 50, 15\}\,,\\
          \text{min\_torque\_lim}=&\{-50, -50, -50, -50, -15\}\,,
\end{align*}
to ensure our simulation is representative of the physical robot limitations. For full parameter details, we encourage the reader to check our implementation \href{https://github.com/lukebhan/NeuralOperatorPredictorFeedback}{Github}
(\url{https://github.com/lukebhan/NeuralOperatorPredictorFeedback}.


\vspace{12pt}
\noindent \textbf{\underline{Dataset generation:}} For data generation, we present all of the data generation parameters in Table \ref{tab:dataGen}. We use the successive approximation solved to a tolerance of $1 \times 10^{-7}$ for generating the ground truth data which we parallelize across $10$ CPU threads taking approximately $3$ hours. Additionally, we note there are a variety of approaches to building sample trajectories of the system. Explicitly, we considered three approaches:
(a) Sample trajectories according to a uniformly varying initial condition;
(b) Sample trajectories with the same initial condition, but inject noise into the numerical predictor akin to the model making a ``wrong'' prediction; 
(c) Following the work of \cite{ross2011reductionimitationlearningstructured}, we build a base dataset using (a) or (b) and then run a trained model online appending all trajectories of the trained model for multiple steps. 
We found that options (a) or (b) work with (c) requiring significant finetuning for the appended dataset. Thus, for the results in Section \ref{sec:numerical_results}, we used a dataset generated with the initial condition $(\text{joint\_lim\_max} + \text{join\_lim\_min})/2$ and with uniform noise ranging from $-0.05$ to $0.05$ added to the final predictor output at each timestep. Lastly, to test our model, we simulated $25$ trajectories with initial conditions perturbed by noise of the form:
\begin{align} \label{eq:sampling_init_cond}
    X(0) \sim (\text{joint\_lim\_max} + \text{join\_lim\_min})/2 + \text{Uniform(-0.05, 0.05)\,. }
\end{align}
For reference, we present the exact parameters for dataset generation, along with an example of a generated trajectory below:

\begin{figure}[H]
\begin{minipage}{0.59\linewidth}
\resizebox{\textwidth}{!}{%
\label{tab:dataGen}
\begin{tabular}{lc}
\hline
\multicolumn{2}{c}{Dataset Generation}                                                              \\ \hline
\multicolumn{1}{l|}{Trajectories}                                                          & 2660    \\
\multicolumn{1}{l|}{\begin{tabular}[c]{@{}l@{}}Samples per\\ trajectory\end{tabular}}      & 94   \\
\multicolumn{1}{l|}{Total samples}                                                         & 250000 \\
\multicolumn{1}{l|}{\begin{tabular}[c]{@{}l@{}}Temporal\\ discretization (s)\end{tabular}} & 0.1                               \\
\multicolumn{1}{l|}{Delay (seconds)}                                                       & $0.5$  \\
\multicolumn{1}{l|}{\begin{tabular}[c]{@{}l@{}}Initial condition \\ sampling\end{tabular}}       & \makecell{$(\text{joint\_lim\_max} + \text{join\_lim\_min})/2$}  \\
\multicolumn{1}{l|}{\begin{tabular}[c]{@{}l@{}}Trajectory length (s)\end{tabular}} & 10      \\ \hline
\end{tabular}
}
\end{minipage}
\begin{minipage}{0.4\linewidth}
    \includegraphics[width=\textwidth]{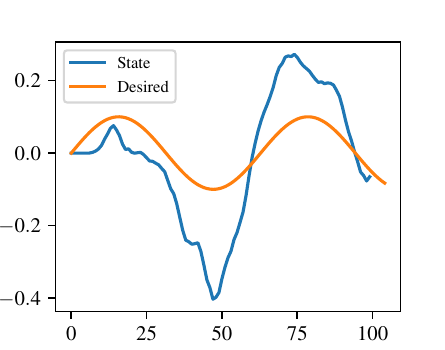}
\end{minipage}
\caption{In the left table, we provide all the dataset generation parameters for the experiments in Section \ref{sec:numerical_results} and on the right figure, we provide an example trajectory using the noise injection as described in (b) above.}
\end{figure}

\noindent \textbf{\underline{Model parameters and training settings:}}
 Note, for each architecture, we perform a hyperparameter search based on the relative $L_2$ test error to obtain the best configurations. We terminate the training once the test error stabilizes and present the full hyperparameters for training in Table \ref{tab:hyperParams} below. We find that the FNO based methods are much easier to train and that the extra parameterization of the spatial-temporal neural operators (FNO+GRU, DeepONet+GRU) does not improve the result, but leads to extensive overfitting for our particular dynamical system. This is perhaps due to the delay $D=0.5$s and an interesting future direction will be testing on different delays.  
\begin{table}[H]
\centering
\caption{Model and training settings. We use the popular optimizer AdamW from \cite{loshchilov2017decoupled}. We use the exponential learning rate scheduler with a varying decay rate.}
\label{tab:hyperParams}
\begin{tabular}{l|l|l|l|l|l}
\hline
Architecture & Epochs & Batch Size & Learning rate & Weight decay & Learning rate scheduler \\ \hline
FNO          & 300    & 512        & 0.005         & 0            & 0.99                    \\
DeepONet     & 500    & 64         & 0.0008        & 0.0001       & 0.999                   \\
FNO+GRU      & 300    & 512        & 0.005         & 0            & 0.99                    \\
DeepONet+GRU & 500    & 64         & 0.0002        & 0            & 0.995 \\          
\hline
\end{tabular}
\end{table}

\noindent \textbf{\underline{Testing the NeuralOP predictor performance in feedback loops:}}
To evaluate our NeuralOP predictor feedback performance in closed-loop, we consider starting the system from initial conditions sampled in \eqref{eq:sampling_init_cond}. Besides the testing case provided in Figure \ref{fig:baxter} of the main paper, we provide an example of the ``worse-case'' performance of the NeuralOP predictor in Figure \ref{fig:worst-case} below. We see that despite the NeuralOP approximation error is larger in this example, it stays within an $\epsilon$-ball around the target trajectory, which obeys with the $\epsilon$-practical stability estimate given in Theorem \ref{thm:main-result}. 
\begin{figure}[H]
    \centering
    \vspace{-10pt}
    \includegraphics[width=0.85\textwidth]{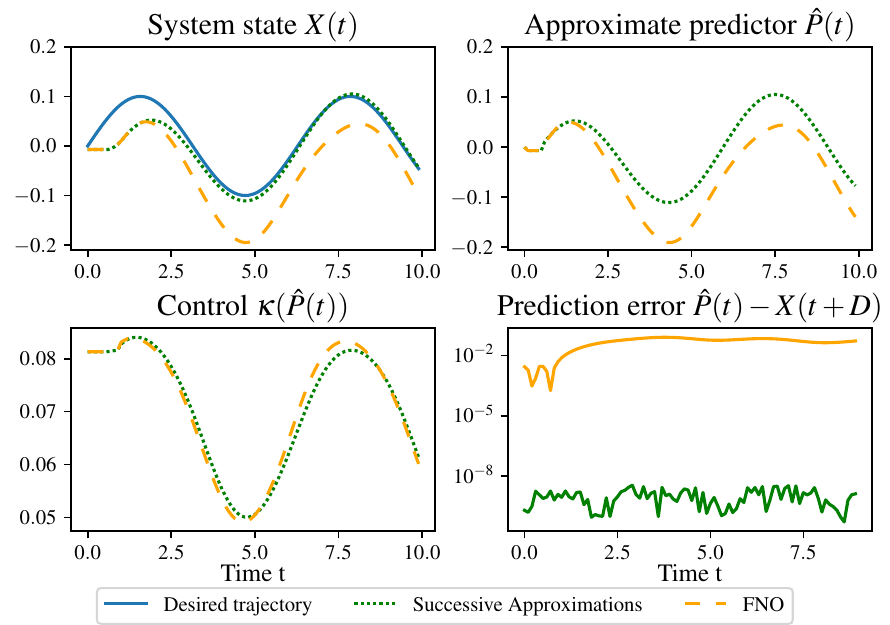}
    \vspace{-10pt}
    \caption{Worse case NeuralOP predictor from the $25$ trajectories in the feedback loop.}
    \label{fig:worst-case}
\end{figure}

\subsection{Comparison of computation time under various discretization sizes and delays} \label{sec:delay_scalability}
As mentioned in the Section \ref{sec:numerical_results}, NeuralOP approximated predictor scales well with respect to the discretization size and delay length compared to the numerical predictor. In Table \ref{tab:numerical}, we quantify the exact improvement averaged over 1000 instances. As step size reduces and delay grows, the acceleration of the NeuralOP approach grows attaining speedups on the magnitude of $420\times$. 
\begin{table}[H]
\vspace{-10pt}
\caption{Comparison of wall-clock time (Nvidia 4060, ms) between different NeuralOP models and the one iteration of the numerical approach for different delay lengths and step sizes. }
\label{tab:numerical}
\begin{tabular}{l|ccc|ccc|ccc}
\hline
\multicolumn{1}{c|}{\textbf{Delay (s)}}            & \multicolumn{3}{c|}{0.1}                                                                                           & \multicolumn{3}{c|}{0.5}                                                                                           & \multicolumn{3}{c}{1}                                                                                              \\ \hline
\multicolumn{1}{c|}{\textbf{Step size $\delta t$}} & 0.1                                  & 0.05                                 & 0.01                                 & 0.1                                  & 0.05                                 & 0.01                                 & 0.1                                  & 0.05                                 & 0.01                                 \\ \hline
Numerical (ms)                                     & 1.78                                 & 3.68                                 & 18.09                                & 8.89                                 & 18.4                                 & 90.74                                & 18.05                                & 35.77                                & 185.11                               \\ \hline
DeepONet                                           & 1.69                                 & 1.34                                 & 1.27                                 & 1.32                                 & 1.26                                 & 1.36                                 & 1.29                                 & 1.32                                 & 1.36                                 \\
FNO                                                & 0.57                                 & {\color[HTML]{009901} \textbf{0.32}} & {\color[HTML]{009901} \textbf{0.32}} & {\color[HTML]{009901} \textbf{0.31}} & {\color[HTML]{009901} \textbf{0.32}} & {\color[HTML]{009901} \textbf{0.34}} & {\color[HTML]{009901} \textbf{0.32}} & {\color[HTML]{009901} \textbf{0.32}} & {\color[HTML]{009901} \textbf{0.44}} \\
FNO+GRU                                            & 1.36                                 & 1.48                                 & 1.43                                 & 1.42                                 & 1.42                                 & 1.57                                 & 1.43                                 & 1.49                                 & 1.68                                 \\
DeepONet+GRU                                       & {\color[HTML]{009901} \textbf{0.45}} & 0.47                                 & 0.46                                 & 0.45                                 & 0.46                                 & 0.54                                 & 0.46                                 & 0.48                                 & 0.63   \\
\hline
\end{tabular}
\vspace{-10pt}
\end{table}

\end{document}